\documentclass [aps,twocolumn,superscriptaddress,altaffilletter,lengthcheck,tightenlines,showpacs,showkeys]{revtex4}
\usepackage[dvipdf]{epsfig}

\newcommand{\ben}{\begin{eqnarray}}
\newcommand{\een}{\end{eqnarray}}
\newcommand{\be}{\begin{equation}}
\newcommand{\ee}{\end{equation}}
\newcommand{\ba}{\begin{eqnarray}}
\newcommand{\ea}{\end{eqnarray}}

\begin{document}


\title{ Wormholes and solitonic shells in five-dimensional DGP theory}


\author{Mart\'{\i}n G. Richarte}\email{martin@df.uba.ar}
\address{ Departamento de F\'{\i}sica, Facultad de Ciencias Exactas y
Naturales,  Universidad de Buenos Aires, Ciudad
Universitaria, Pabell\'on I, 1428 Buenos Aires, Argentina}

\begin{abstract}
We build five-dimensional spherically symmetric wormholes within the DGP theory. We calculate the energy localized on the shell, and we find that the wormholes could be supported by matter not violating the energy conditions. We also show that  solitonic  shells characterized by zero pressure and zero energy can exist; thereafter  we make some  observations regarding their  dynamic on the  phase plane. In addition, we concentrate on the mechanical stability of  wormholes under radial perturbation preserving the original spherical symmetry.  In order to do that, we consider  linearized perturbations around  static solutions. We obtain that for certain values  of the mass $\mu$ and crossover scale $r_{c}$ stable wormholes exist  with very small values  of  squared speed sound. Unlike the case of Einstein's gravity, this type of wormholes fulfills the energy conditions. Finally, we show that the gravitational field associated with  these wormhole configurations is attractive for $\mu>0$. 
\end{abstract}

\maketitle

\section{Introduction}
 Traversable Lorentzian wormholes  \cite{motho,visser} are topologically non trivial solutions of the equations of gravity which would imply a  connection between two regions of the same universe, or of two universes, by a traversable  throat.  In the case that such geometries actually exist they could show some interesting peculiarities as, for example, the possibility of using them for  time travel \cite{morris,novikov}. A basic difficulty with wormholes is that the  flare-out condition \cite{hovis1} to be  satisfied at the throat  requires the presence of matter which violates the energy conditions (``exotic matter'') \cite{motho,visser,hovis1,hovis2}. It was recently shown \cite{viskardad}, however, that the amount of exotic matter necessary for supporting a wormhole geometry can be made infinitesimally small. Thus, in subsequent works special attention has been devoted to quantifying the amount of exotic matter  \cite{bavis,nandi1}, and  this measure of the exoticity has been pointed to as an indicator of the physical viability of a traversable wormhole \cite{nandi2}.

A central aspect of any solution of the equations of gravitation is its mechanical stability. The stability of wormholes  has been thoroughly studied for the case of small perturbations preserving the original symmetry of the configurations. In particular, Poisson and Visser \cite{poisson} developed a straightforward approach for analyzing this aspect for thin-shell wormholes; that is, those which are mathematically constructed by cutting and pasting two manifolds to obtain a new manifold \cite{mvis1, mvis2}. In these wormholes the associated  supporting matter is located on a shell placed at the joining surface; so the theoretical tools for treating them is the Darmois-Israel formalism, which leads to the Lanczos equations \cite{daris, mus}. The solution of the Lanczos equations gives the dynamical evolution of the wormhole once an equation of state for the matter on the shell is provided.  Such a procedure has been subsequently followed  to study the stability of more general spherically  symmetric configurations (see, for example, Refs. \cite{ eirom1, eirom2, eirom3, eirom4, eirom5, eirom6, eirom7,
eirom8, eirom9, eirom10}). Moreover, the junction conditions were also used to construct plane symmetric thin-shell wormholes with cosmological constant \cite{lemos1, lemos2}.

Wormholes in theories beyond Einstein framework  have gained a lot of interest in the last years  because they  seem to possess some curious  properties regarding the kind of matter that could support them. A few  examples of these alternatives theories  are the Einstein-Gauss-Bonnet picture \cite{mi1, whegb1a, whegb1b, whegb2}, scalar-tensor theories \cite{mi2, whbdicke1, whbdicke2, whbdicke3}, $F(R)$theory, or massive gravity \cite{whfrotros1, whfrotros2, whfrotros3, whfrotros4, whfrotros5}. In particular, for the Einstein-Gauss-Bonnet theory, it was shown that static thin-shell wormholes could be supported by ordinary matter respecting the energy conditions\cite{mi1}. Moroever, $C^2$-type womholes with the latter property can also exist once the nonlinear Gauss-Bonnet term is included in the field equations \cite{whegb1a, whegb1b}. Of course, this feature is not only exclusive of the Gauss-Bonnet paradigms; the Brans-Dicke gravity is  another set up where the thin-shell wormholes fulfill weak and null  energy conditions \cite{mi2}. 

In addition,  a new type of gravitational model was widely studied in the context of cosmology as well as particle physics, the so-called Dvali, Gabadadze, and Porrati (DGP) theory. It predicts deviations from the standard 4D gravity over large distances. The transition between four- and higher-dimensional gravitational potentials in the
DGP model arises because of the presence of both the brane and the bulk Einstein-Hilbert (E-H) terms in the action \cite{dgp1, dgp2}.
 Cosmological considerations of the DGP model were first discussed in \cite{ace1,ace2}. where it was
shown that in a Minkowski bulk spacetime we can obtain self-accelerating solutions. In the original
DGP model it is known that 4D general relativity (GR) is not recovered at the linearized level. However, some
authors have shown that at short distances we can recover the 4D general relativity in a spherically
symmetric configuration (see, for example, \cite{tanaka}).

It is worth mentioning that an interesting feature of the original DGP model is the existence of
ghostlike excitations \cite{exic1, exic2, jc1a, jc1b}. Further,  the viability of the self-accelerating cosmological solution  in the DGP gravity was carefully studied in \cite{dimitri}. For a comprehensive review of the existence of 4D ghosts on the self-accelerating branch of solutions in DGP models, see \cite{jc2a, jc2b}.

A common feature among alternative theories is that the junction conditions for the thin-shell wormholes  are modified considerably, adding new types of geometrical objects besides the usual extrinsic curvature. The contributions from the curvature tensors, theoretically, seem to allow the existence of  wormholes supported by ordinary matter. For all these reasons, we consider that  the construction of  wormholes within DGP gravity deserves to be examined in detail  to conclude whether they could  fulfill or not the energy conditions.

Another consequence of the nonlinearity introduced by the DGP theory  is related to  the way in which the stability analysis is carried out for the dynamic case; that is, in this context it is not completely clear how to obtain the stability zones.

The aim of the present paper is twofold. On the one hand, we explore the existence of five-dimensional wormholes within the DGP gravity theory. Our research is focused on configurations supported by nonexotic matter which satisfies the energy conditions. Then, we show the existence of solitonic vacuum shells  and make some comment about their dynamic. On the other hand, our goal is to perform  a study  of the linear  stability of  wormholes  preserving  the original symmetry. We only examine  configurations supported by ordinary matter. Moreover, we shall show that there exist  stable wormholes with  squared speed sound  within the range $0\leq v^{2}_{~s}\leq 0.1$ indicating that the matter located at the throat of the wormhole could be nonrelativistic.

\section{Five-dimensional bulk solution}
We start from the action for the DGP theory in five-dimensional manifold ${\cal{M}}_5$  with four-dimensional boundary $\partial{\cal{M}}_5=\Sigma$ (cf. \cite{jc1a, jc1b, jc2a, jc2b, jc3a, jc3b}),
\begin{eqnarray*}
\label{dgpa}
S&=&2M^{3}_{5}\int_{{\cal M}_{5}} d^{5}{x}\sqrt{-g}R(g_{\mu\nu}) + 2M^{2}_{4}\int_{\Sigma} d^{4}{x}\sqrt{-\gamma}{\cal R}(\gamma_{ab}) \\ 
&+&\int_{\Sigma} d^{4}{x}\sqrt{-\gamma}\Big(-4M^{3}_{5}{\cal{K}}(\gamma_{ab}) + {\cal{L}}_{m}\Big),\\
\end{eqnarray*}
where $g_{\mu\nu}$ is the five-dimensional metric, $\gamma_{ab}$ is the four-dimensional induced metric on the boundary $\Sigma$, and $\cal{K}$ is the trace of extrinsic curvature. 
The extra term in the boundary introduces a mass scale $m_{c}=2M^{3}_{5}/M^{2}_{4}=r^{-1}_{c}$; that is, the model has one adjustable parameter, namely, $m_c$ which determines a scale that separates two different regimes of the theory. For distances much smaller than $m^{-1}_{c}$  one would expect the solutions to be well approximated by general relativity and the modifications to appear at larger distances. This is indeed the case for distributions of matter and radiation which are homogeneous and isotropic
at scales $\gtrsim r_{c}$. Typically, $m_{c}\sim 10.42 \mbox{GeV}$, so it sets the distance/time scale $r_{c}=m^{-1}_{c}$ at which the Newtonian potential significantly deviates from the conventional one (cf.\cite{foot1a, foot1b}).

It is a well-known fact that the DGP scheme is a five-dimensional model where gravity propagates throughout an infinite bulk, and matter fields in ${\cal{L}}_{m}$ are confined to a 4-dimensional boundary. The action for gravity at lowest order in the derivate expansion is a bulk Einstein-Hilbert term and a boundary one, generically with two different Planck masses $M_5$, $M_4$, plus a suitable Gibbons-Hawking term.
In the bulk the DGP  equations are the Einstein ones in vacuum: $G^{(5)}_{\mu\nu}=0$. Then in this case,  Birkhoff's theorem forces the bulk metric to be static, and of the  Schwarzschild form:
\begin{eqnarray}
\label{metric1} 
ds^{2}&=& -f(r) dt^{2} + \frac{1}{f(r)}dr^{2} + r^{2}d\Omega^{2}_{3},
\\
\label{metric2} 
f(r)&=&1- \frac{\mu}{r^2}
\end{eqnarray}
where the parameter $\mu$ is related to the five-dimensional Arnowitt-Deser-Misner (ADM) mass, $M_{_{ADM}}=3\pi^2\mu M^3_{5}$. The above spacetime has only one horizon placed at $r_{+}=\sqrt{\mu}$ with $\mu>0$. Besides, when $\mu<0$ the manifold only presents a naked singularity at the origin $r=0$. Now, we are going to make some important remarks about this solution: (i) notice that the $r$ does not measure the 4D distance from the origin because it has a spatial component in the extra dimension labeled as $y$. Therefore, one can write  the 5D distance as  $r^{2}:=r^{2}_{4D}+y^{2}$; (ii) second, at  large enough distance the five-dimensional Schwarzschild solution naturally appears in DGP gravity with correct boundary conditions. However, the behavior of the bulk solution and the corresponding black holes localized on a brane within DGP gravity  are quite different in some regimes \footnote{
 A more  careful analysis indicates that when  compact static sources of  mass M and radius $r_{0}$ are taken into account, such that $r_{M}<r_{0} << r_{c}$ ($r_M= 2MG$ is the Schwarzschild radius), a new scale, a combination of $r_c$ and $r_g$, emerges (the so-called Vainshtein scale): $r_{*}=(r^{2}_{c}r_{M})^{1/3}$.
Below this scale the predictions of the theory are in good agreement with the GR results and above it they deviate considerably (cf. \cite{foot1a}, \cite{foot1b}). }. Further, regarding the four-dimensional point of view (localized on a DGP brane) in \cite{foot2} it was found that the theory admits a 4D anti-de Sitter (AdS-)Schwarzschild solution; there the four-dimensional cosmological constant is given by $\Lambda_{4D}=-3m^{2}_{~c}<0$. In that work, the authors also showed that in the  limit $r_{4D}>>m^{-1}_{~c}$ the 
5D Schwarzschild solution of radius $r_{g}$ emerges, wheareas for $r_{4D}<<m^{-1}_{~c}$ the metric can be accommodated in 4D Schwarzschild geometry. Interestingly enough,  there is an interpolating solution between these two regimes (regular branch) together with a second solution that becomes 5D de Sitter-Schwarzschild at a large distance (accelerated branch)(cf.\cite{foot2}).


\section{Wormholes in DGP theory}
\subsection{Thin-shell construction}
Employing the metric Eqs.(\ref{metric1}-\ref{metric2}) we build a spherically thin-shell wormhole in DGP theory. We take two copies of the spacetime and remove from each manifold the five-dimensional regions described by 
\begin{equation}
{\cal M}_{\pm}=\left\{x/r_{\pm}\leq a,a>r_{h}\right\}.
\end{equation} 
The resulting manifolds have boundaries given by the timelike hypersurfaces 
\begin{equation}
\Sigma_{\pm}=\left\{x/r_{\pm} = a,a>r_{h}\right\}.
\end{equation}
Then we identify these two timelike hypersurfaces to obtain a geodesically complete new manifold ${\cal {M}}={\cal {M}}^{+}\cup {\cal {M}}^{-}$. We take values of $a$ large enough to avoid the presence of singularities and horizons in the case that the geometry (\ref{metric2}) has any of them. The manifold $\cal M$ represents a wormhole  with a throat placed at the surface  $r=a$, where the matter supporting the configuration is located. This manifold is constituted by two regions which are  asymptotically flat (see Fig. 1).
\begin{figure}[!h]
\begin{center}
\includegraphics[height=9cm, width=8cm]{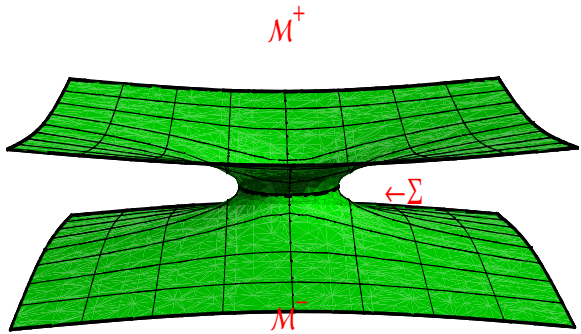}
\caption{We show the wormhole geometry obtained after performing the cut and paste procedure. The shell on $\Sigma$ is located at the throat radius $r=a$.}
\end{center}
\end{figure}
The wormholes throat $\Sigma$ is a synchronous timelike hypersurface, where we define locally a chart with coordinates $\xi^{a}=(\tau,\chi,\theta,\phi)$, with $\tau$ the proper time on the shell.  Though  we shall first focus in static configurations, in the subsequent  we could allow the radius of the throat to be a function of the proper time for studying the dynamics evolution of the wormholes, and then in general we have that the  boundary hypersurface reads
\begin{equation}
\Sigma: {\cal H}(r,\tau)=r-a(\tau)=0.
\end{equation}
 It is important to remark that the geometry remains static outside the throat, regardless of the fact that the radius $a(\tau)$ can vary with time, so no gravitational waves are present. This is naturally guaranteed because the Birkhoff theorem holds for the original manifold.

Our starting point is to list the main geometric objects which shall appear in the junction condition associated with the field equation for $\Sigma$.
The extrinsic curvature,namely,  ${\cal K}_{ab}$, associated with the two sides of the shell is defined as follows:

\begin{equation}
{{\cal K}}^{\pm}_{ab}=-n^{\pm}_{\kappa}\left(\frac{\partial^{2}X^{\kappa}}{\partial\xi^{a}\partial\xi^{b}}+\Gamma^{\kappa}_{\mu\nu}\frac{\partial X^{\mu}}{\partial\xi^{a}}\frac{\partial X^{\nu}}{\partial\xi^{b}}\right)_{r=a},
\end{equation} 
where $n^{\pm}_{\kappa}$ are the units normal($n_{\kappa}n^{\kappa}=1$) to the surface $\Sigma$ in ${\cal M}$:
\begin{equation} 
n^{\pm}_{\kappa}=\pm\left| g^{\mu\nu}\frac{\partial {\cal H}}{\partial X^{\mu}}\frac{\partial {\cal H}}{\partial X^{\nu}} \right|\frac{\partial {\cal H}}{\partial X^{\kappa}}
\end{equation}

The  field equations projected on the shell $\Sigma$ are the generalized junction (or Darmois--Israel) conditions \cite{jc1a, jc1b, jc2a, jc2b} 
\begin{equation}
\label{jc}
r_{c}\Big({\cal R}_{ab}-\frac{1}{2}\gamma_{ab}{\cal R}\Big)-2\Big(\left\langle {\cal K}_{ab}-{\cal K}\gamma_{ab}\right\rangle\Big)=\frac{{\cal S}_{ab}}{8M^3_{5}},
\end{equation}
where the bracket $\left\langle .\right\rangle$ stands for  the jump of a given quantity across the  hypersurface $\Sigma$ and  $\gamma_{ab}$ is the induced metric on $\Sigma$.
Notice that the first term in (\ref{jc}) is not enclosed with the brackets because this contribution comes from the  four-dimensional E-H term in the DGP action (\ref{dgpa}) which already lives in the boundary so it does not need to be projected on $\Sigma$. By taking the limit  $r_{c}\rightarrow 0$ we recover the standard Darmois-Israel junction condition found in \cite{daris}.

Now, let us calculate some quantities that we shall need later. The mixed components of the four-dimensional Einstein tensor are given by
\begin{eqnarray}
\label{ET} 
{\cal G}^{0}_{~0}&=&-3\Big(\frac{{\dot{a}}^{2}}{a^2} + \frac{1}{a^2}\Big),
\\
\label{EE} 
{\cal G}^{i}_{~j}&=&-\Big(\frac{1}{a^2}+ \frac{{\dot{a}}^{2}}{a^2}+ 2\frac{\ddot{a}}{a}\Big)\delta^{i}_{~j}
\end{eqnarray}
where the dot means derivate with respect to the proper time on $\Sigma$. The extrinsic curvature components read
\begin{eqnarray}
\label{KT} 
\left\langle {\cal K}^{0}_{~0} \right\rangle&=&\frac{2\ddot{a} + f'(a)}{\sqrt{f(a)+{\dot{a}}^{2}}},
\\
\label{KE} 
\left\langle {\cal K}^{i}_{~j}\right\rangle&=&\frac{2}{a}\sqrt{f(a)+{\dot{a}}^{2}}~\delta^{i}_{~j}
\end{eqnarray}
where the prime indicates the derivates with respect to $a$. The  most general form of the stress-energy tensor on shell compatible with the symmetries is 
 \begin{equation}
 \label{tem}
 {\cal S}^{a}_{~b}=~\mbox{diag}~(-\sigma, p ~\delta^{i}_{~j})
\end{equation}
where $\sigma$ is the energy density and $p$ is the pressure. Replacing Eqs(\ref{EE}-\ref{tem}) into the DGP junction condition(\ref{jc}) we obtain that the energy density and the pressure can be recast as
\begin{eqnarray}
\label{sigma} 
\frac{\sigma}{8M^3_{5}}&=&3r_{c}\Big(\frac{{\dot{a}}^{2}}{a^2} + \frac{1}{a^2}\Big)-\frac{12}{a}\sqrt{f(a)+{\dot{a}}^{2}},
\\
\label{pe} 
\frac{p}{8M^3_{5}}&=&-r_{c}\Big(\frac{{\dot{a}}^{2}}{a^2} + \frac{1}{a^2}+ \frac{2\ddot{a}}{a}\Big)+\frac{8}{a}\sqrt{f(a)+{\dot{a}}^{2}}\\ 
&+&2\frac{2\ddot{a}+f'}{\sqrt{f(a)+{\dot{a}}^{2}}}.
\end{eqnarray}
where the DGP contributions are encoded in the $r_{c}$ factor of the above equations. If we take  $r_{c}\rightarrow 0$ in both Eqs.(\ref{sigma}) and (\ref{pe}) we recover  the expression for the energy density $\sigma$ and the pressure $p$ found in \cite{mi1}, ignoring the Gauss-Bonnet contribution.

In order to carry on let us comment that we still have the  usual energy conservation,$\nabla_{a}{\cal S}^{ab}=0$ by virtue of $\nabla^{a}({\cal K}_{ab}-\gamma_{ab}{\cal K})=0$, coming from the momentum constraint implicit in the five-dimensional Einstein equations. Further it is easy to see from $\sigma$  and $p$ that the energy conservation equation is fulfilled:
\begin{equation}
\frac{d(a^{3}\sigma)}{d\tau}+p\frac{da^{3}}{d\tau}=0,\label{conserva}
\end{equation}
The first term in Eq. (\ref{conserva}) represents the
internal energy change of the shell and the second the work by internal forces of the
shell. The dynamical evolution of the wormhole throat is governed by the generalized Lanczos equations and to close the system we must supply an equation of state $p = p(\sigma)$ that relates $p$ and $\sigma$. Notice that the reason why one obtains exact conservation, i.e., no energy flow to the bulk, is that the normal-tangential components of the stress tensor in the bulk are the same on both sides of the junction hypersurface.

\section{Matter supporting the wormholes}
Recently, classical solutions  within  the DGP model were found when the stress-energy tensor on the brane satisfies the dominant energy condition, yet the brane has negative energy from the bulk point of view (see \cite{jc1a, jc1b}). Within this frame, the study of superluminal propagation  indicates  that superluminosity occurs whenever the stress tensor on the shell is a pure cosmological constant, irrespective of the value  of the shell density (cf.\cite{jc1a, jc1b}). All these elements are  good reasons to consider a careful discussion about the nature of matter supporting wormholes in the DGP model.
Moreover, motivated by the results within Einstein-Gauss–Bonnet gravity (i.e. with $R^{2}$-like terms) in \cite{mi2},
here we evaluate the amount of exotic matter and the energy conditions, following the approach
presented above where the four-dimensional EH term generalizes the standard junction, adding  a few  geometrical terms, which indeed represents  the Einstein tensor projected on the shell.  Consequently, coming the DGP contribution from the curvature tensor, the next approach is clearly the most suitable to give a precise meaning to the characterization of matter supporting the wormhole.

The \emph{weak} energy condition (WEC) states that for any timelike vector $U^{\xi}$ it must be $T_{\xi\eta}U^{\xi}U^{\eta}\geq 0$;
the WEC also implies, by continuity, the \emph{null} energy condition (NEC), which means that for any null
vector $k^{\xi}$ it must be $T_{\xi\eta}k^{\xi}k^{\eta}\geq 0$ . In an orthonormal basis the WEC reads $\rho\geq 0$, $\rho + p_{l}\geq 0$ $\forall ~ l$  while the NEC takes the form $\rho + p_{l}\geq 0$ $\forall ~ l$.  Besides, the \emph{strong} energy condition (SEC) states that $\rho + p_{l}\geq 0$ $\forall ~ l$, and $\rho + 3p_{l}\geq 0$ $\forall ~ l$.

In the case of thin-shell wormholes the radial pressure $p_r$ is zero,  within Einstein gravity, and the surface energy density must fulfill $\sigma < 0$, so that both energy conditions would be violated. The sign of $\sigma+p_{t}$ where $p_t$ is the transverse pressure is not fixed, but it depends
on the values of the parameters of the system. In what follows we restrict to static configurations. The surface energy density $\sigma_{0}$ and the
transverse pressure $p_{0}$ for a static configuration ($a = a_0$, $\dot{a}=0$, and $\ddot{a}=0$) are given by

\begin{eqnarray}
\label{sigo} 
\frac{\sigma_{0}}{8M^3_{5}}&=& \frac{3r_{c}}{a^{2}_{0}}-\frac{12}{a_{0}}\sqrt{f(a_{0})},
\\
\label{po} 
\frac{p_{0}}{8M^3_{5}}&=&- \frac{r_{c}}{a^{2}_{0}}+\frac{8}{a_{0}}\sqrt{f(a_{0})}+ 2\frac{f'(a_{0})}{\sqrt{f(a_{0})}}.
\end{eqnarray}
Now the sign of the surface energy density  as well as the pressure is, in principle, not fixed. The most usual choice  for quantifying the  amount of exotic matter in a Lorentzian wormhole is the integral \cite{nandi1}:
\begin{equation}
\Omega= \int (\rho + p_{r})\sqrt{-g_{5}}\,d^{4}x.
\end{equation}
We can introduce a new radial coordinate $R=\pm(r-a_{0})$ with $\pm$ corresponding to each side of the shell. Then,  
because  in our construction the energy density is located on the surface, we can also write $\rho=\delta(R)\,\sigma_{0}$, and because the shell does not exert radial pressure  the amount of exotic matter reads
\begin{equation}
\Omega=\int\limits^{2\pi}_{0} \int\limits^{\pi}_{0}\int\limits^{\pi}_{0}\int\limits^{+\infty}_{-\infty}\delta(R)\,\sigma_{0} \sqrt{-g_{5}}\, d R\, d\xi\,d\theta\,d\phi\ =2\pi^{2}  a_{0}^3 \sigma_{0}.
\end{equation}
Replacing the explicit form of $\sigma_{0}$ and $g_{5}$, we obtain the exotic matter amount as a function of the parameters  that characterize the configurations:
\begin{equation}
\label{ome1}
\Omega=16M^3_{5}\pi^{2}\Big(~3r_{c}~ a_{0}-12a^{2}_{0}\sqrt{ f(a_{0})}~\Big).
\end{equation}
where $f$ is  given by the bulk solution. For $r_{c}\rightarrow 0$ we obtain the exotic amount for Schwarzschild geometries  as if it were calculated  with the standard junction conditions.
Far away from the general relativity limit  we now find that there exist positive contributions to  $\sigma_{0}$; these come from the different
signs in the expression (\ref{ome1}) for the surface energy density, because it is proportional to $\sigma_{0}$. We stress
that this would not be possible if the standard Darmois–-Israel formalism was applied, treating the
DGP contribution as an effective energy-momentum tensor, because this leads to $\sigma_{0}\propto - \sqrt{ f(a_{0})}/a_{0}$. Now, once the explicit form of the function $f(a_{0})$ is introduced in Eq.($\ref{ome1}$), we focus on  what are the conditions  that lead to wormholes with $\sigma_{0}>0$ or $\Omega>0$. Then, it can be proved that wormholes with a non-negative surface density located at the shell are  allowable when the following inequalities are simultaneously satisfied: 

\begin{eqnarray}
\label{snula2} 
\frac{r_{c}}{a^{2}_{0}}-\frac{4}{a_{0}}\Big(1-\frac{\mu}{a^{2}_{0}}~\Big)^{1/2}&>&0,
\\
\label{fueho1} 
a^2_{0}-\mu&>0&,
\end{eqnarray}
so it is always possible  to choose $a_{0}$ such that  the existence  of thin-shell wormholes is compatible with positive  surface energy density (see Fig2.); more precisely its radius  must belong to  the interval given as
\begin{equation} 
 \label{int1a}
 \sqrt{\mu}<a_{0}\leq \big(\mu + \frac{r^{2}_{c}}{16}\big)^{\frac{1}{2}}
\end{equation}
Notice that the $r_{c}$term is essential to have positive energy density; as one would expect, in the limit $r_{c} \longrightarrow 0$, this possibility completely vanishes. Despite that in most of our discussions we are going to keep $r_{c}$ and $\mu$ as  free parameters; we wish to discuss some order of magnitude for $r^{2}_{c}/\mu $, so that from Eq. (\ref{int1a}) one can immediately see whether there is a fine-tuning or not. Using constraints from type 1A supernovae \cite{rc} it turns out to be that the best fit for the crossover scale is $r_{c}= (1.21 \pm 0.09 )H^{-1}_{~0}$, where $H_{0}$ is today's Hubble scale. Taking $H_{0}=70\mbox{km} \mbox{s}^{-1}\mbox{Mpc}^{-1}$, it implies that $r_{c}\sim 5\mbox{Gpc}$; just to have an idea of this magnitude it is useful to remember that  the distance to largest structures in the distribution galaxies is $100\mbox{Mpc}$ while the distance to the edge of the visible universe is $14\mbox{Gpc}$. On the other hand, four-dimensional Planck mass is given by $M_{4}=1.22\times10^{19}\mbox{Gev}$ so using the definition of $r_{c}=M^{2}_{~4}/2M^{3}_{~5}$ we obtain that the five-dimensional Planck mass (scale) is $M_{5}\sim 45 \mbox{Mev}$. This value for the fundamental scale  was also found in \cite{pheno1a}; there are other interesting works that show more constraints on what is the value that it should take the scale $M_{5}$(see \cite{pheno1b}, \cite{pheno1c}). In addition, a bound for the scale $M_{5}$ can be also obtained by using the Solar System itself \cite{pheno2} or outer planets of the Solar System\cite{pheno3}. Coming back to our main analyses, we are in a position to estimate in broad terms the ratio $r^{2}_{c}/\mu$. Taking $M_{ADM}\sim M_{\mbox{sun}}$  and $r_{c}\sim 5\mbox{Gpc}$ implies that $r^{2}_{c}/\mu \sim {\cal O}(10^{24})$ or $r^{2}_{~c}>>\mu$; consequently it indicates that there is no fine-tuning at all. Using the latter results we get an estimation of $\mu \sim 4\times10^{59}\mbox{Gev}^{-2}$, where in order to  obtain the $M_{ADM}$ we have assumed that at enough large distance the metric should correspond to asymptotically flat spacetime. For a full treatment about  the role of the boundary conditions for obtaining a numerically black  hole solution within the Randall-Sundrum infinite braneworld see \cite{tana2}. 


\begin{figure}[!h]
\includegraphics[height=8cm, width=7cm]{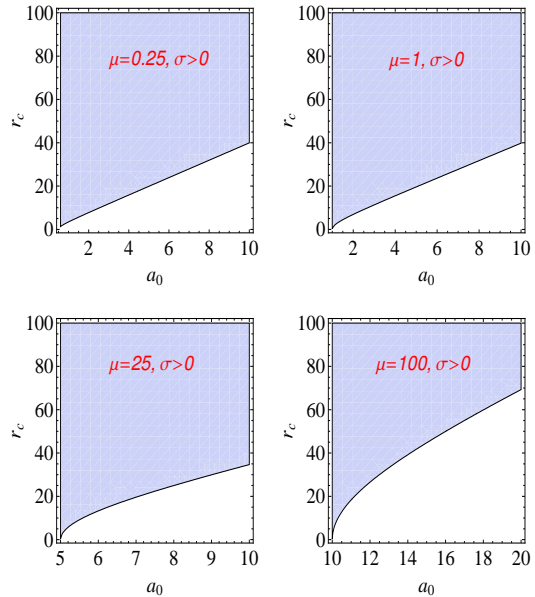}
\caption{We plot the zones in the plane $r_{c}-a_{0}$ where the condition $\sigma_{0}>0$ for several values of $\mu$.}
\end{figure}
Besides, from Eq.(\ref{sigo}) and Eq.(\ref{po}) we have that  the sum of the pressure and energy density  takes the form
\begin{equation} 
 \label{int2}
 \sigma_{0}+ p_{0}=8M^3_{5}\left(\frac{2r_{c}}{a^{2}_{0}}+\frac{2a_{0} f'(a_{0})-4f(a_{0})}{a_{0}\sqrt{f(a_{0})}}\right)
\end{equation}
because the first  term in (\ref{int2}) is positive the sign of $\sigma_{0}+ p_{0}$ depends on the second term, implying that the sum is positive  for $\sqrt{\mu}<a_{0}\leq \sqrt{2\mu}$. Therefore, the remarkable result is that we have a region with $\sigma_{0}\geq0$ and besides $\sigma_{0}+ p_{0}\geq0$ , so the WEC and the NEC are satisfied (see Figs.3 and 4). Additionally, it is easy to corroborate that $\sigma_{0}+ 3p_{0}=12\times8M^3_{5}/(a_{0}\sqrt{f(a_{0})}) $, then  SEC holds in the interval $a_{0} \in (\sqrt{\mu}, \sqrt{2\mu}]$ (see Figs.3 and 4). Thus, by treating the DGP contribution as a geometric object, the generalized junction conditions (\ref{jc}) provide  a clear meaning to the matter in the shell leading to a central finding that in the  DGP gravity the violation of the energy conditions could be avoided and wormholes could be supported by ordinary matter.

\begin{figure}[!h]
\includegraphics[height=7cm, width=9cm]{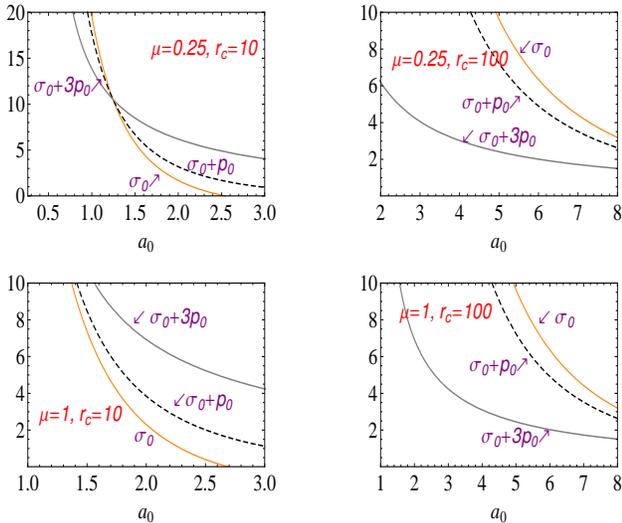}
\caption{We show  $\sigma_{0}$, $\sigma_{0}+p_{0}$ and $\sigma_{0}+3p_{0}$ versus the wormhole radius $a_{0}$ for several values of $(\mu,r_{c})$.}
\end{figure}

\begin{figure}[!h]
\includegraphics[height=7cm, width=9cm]{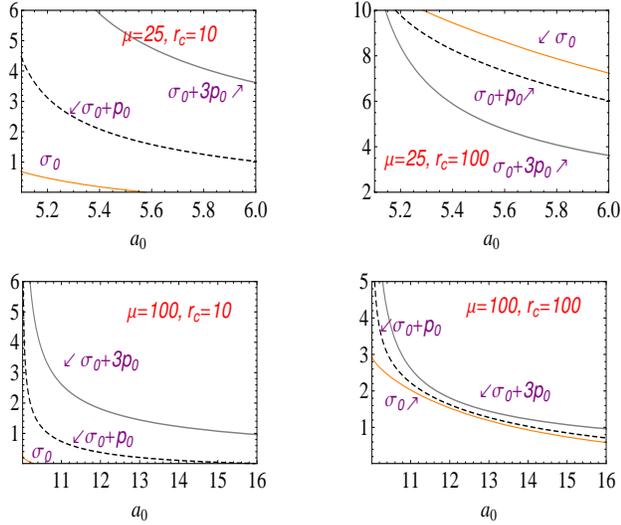}
\caption{We show  $\sigma_{0}$, $\sigma_{0}+p_{0}$ and $\sigma_{0}+3p_{0}$ versus the wormhole radius $a_{0}$ for different values of $(\mu,r_{c})$.}
\end{figure}
However, note that one could choose another route because  Eq.(\ref{jc}) can be formally recast as follows: 
\begin{eqnarray}
-16M^3_{5}\left\langle {\cal K}_{ab}-{\cal K}\gamma_{ab}\right\rangle&=&{\cal S}^{eff}_{ab},
\\
{\cal S}_{ab}-8M^3_{5}r_{c}\Big({\cal R}_{ab}-\frac{1}{2}\gamma_{ab}{\cal R}\Big)&=&{\cal S}^{eff}_{ab}
\end{eqnarray}
although this identification is also possible; physically we would be  treating curvature objects  as an effective source for the junction condition. Moreover, based on the effective energy-momentum tensor approach we  inevitably would obtain that the energy density is  negative definite because the flare-out condition is fulfilled. For a review  of junction conditions within the DGP theory see \cite{jc3a, jc3b} and references therein.

\section{Solitonic wormholes/shells}

Now, we are going to focus on a particular type of wormholes/shells. To be precise we desire to examine if it is possible to have dynamical solitonic wormholes/shells characterized by a zero pressure ($p=0$) and zero energy density ($\sigma=0$).
Unlike the standard Darmois-Israel junction condition, nontrivial  solutions may be possible even when ${\cal S}^{a}_{b}=0$. That is, the extrinsic curvature can be discontinuous across the throat with no matter on the shell to serve a source, turning the discontinuity into a self-supported gravitational system. Of course, these configurations are impossible in the Einstein gravity but not in the Einstein-Gauss-Bonnet gravity (cf. \cite {whegb2}).

For $\dot{a}\neq 0$  Eq.(\ref{conserva}) shows that  if $\sigma=0$ then $p=0$; so we are going to work with the most useful expression which in this case is  given by $\sigma$.  Following the procedure  mentioned in \cite{jc2a}  we shall plot  trajectories in the  phase space spanned by $(\dot{a},a)$. Because  the energy constraint ($\sigma=0$) is invariant under the symmetry $\dot{a}\longleftrightarrow -\dot{a}$  we can work on a two-dimensional plane which is defined as a noncompact domain, namely, ${\cal B}=(0, +\infty) \times (r_{+}, +\infty)$. The curves which represent  the dynamics of solitonic wormholes are obtained by imposing the following conditions:
\begin{eqnarray}
\label{snula} 
r_{c}\big({\dot{a}}^{2} + 1\big)-4\Big(a^2-\mu +a^2{\dot{a}}^{2}\Big)^{\frac{1}{2}}&=&0,
\\
\label{fuerahorionte} 
a^2-\mu&>0&
\end{eqnarray}
such that the first inequality guarantees zero energy, whereas the second one ensures that the wormhole radius is larger than the event horizon.
\begin{figure}[!h]
\label{sp}
\includegraphics[height=11cm, width=9cm]{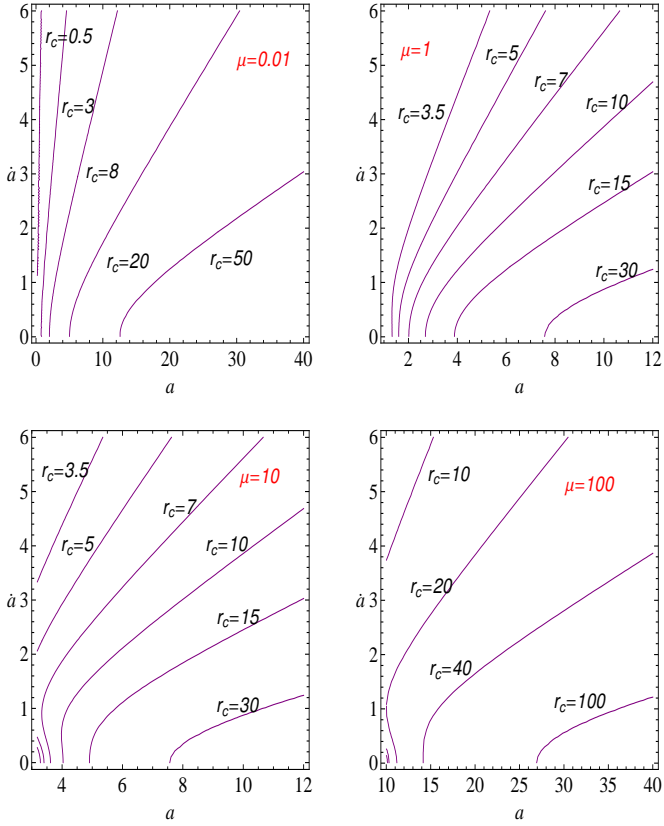}
\caption{We show the trajectories of vacuum shells in the phase plane for $\mu \in [0.01,100]$ and different values of DGP scale $r_{c}$.}
\end{figure}
According to Fig. 5, the  phase diagrams show that for small or large $\mu$ and with a DGP scale covering the interval $[0.5,50]$, the shell velocity is a monotone increasing function for $\dot{a}\in(0,+\infty)$ [or decreasing one  when $\dot{a} \in (-\infty, 0)$]. Notice that the same conclusion is obtained when  the parameter $r_{c}$ takes larger values. In order to see if these types of shells speed up or decelerate we use the zero pressure condition to get a functional relation $\ddot{a}={\cal N}[a,\dot{a}]$ which determines the sign of $\ddot{a}$:
\begin{equation}
\label{signAc}
{\cal N}=\frac{-8af(a)-8a{\dot{a}}^{2}-2a^{2}f'(a)+r_{c}(1+{\dot{a}}^{2})\sqrt{f(a)+{\dot{a}}^{2}}}{2a\Big(2a- r_{c}\sqrt{f(a)+{\dot{a}}^{2}}\Big)}
\end{equation}
For all $\mu$ and $ r_{c}$ considered in this section  we obtain that the kinematic of the shell has four possible types of dynamical evolution. More precisely, the solitonic solution  could suffer an accelerated ($\ddot{a}>0$) or decelerated ($\ddot{a}<0$) expansion ($\dot{a}>0$)  as well as  an accelerated or decelerated contraction ($\dot{a}<0$) regimes.

Unlike the Einstein-Gauss-Bonnet case  studied in \cite{whegb2} it turns out that the existence of solitonic shells in  DGP gravity  does not require the presence of a cosmological constant term in the bulk spacetime.  

In the next section we are going to study the stability of five-dimensional wormholes against homogenous perturbations preserving the original symmetry.

\section{The stability analysis}

In general to obtain the dynamic picture  of the wormholes within the DGP gravity is a very complicated task. As  can be seen from  Eqs. (\ref{sigma}-\ref{pe}) the nonlinear character of these expressions makes   the idea of obtaining exact solutions  very hard to implement. However,  we can follow another route and study the stability of static solutions by linearizing the field equation. 
A physically interesting wormhole geometry should last enough so that its traversability
makes sense. Thus the stability of a given wormhole configuration becomes a central
aspect of its study. Here we shall analyze the stability under small perturbations
preserving the spherical symmetry of the configuration; for this we shall proceed as
\cite{poisson}, \cite{ersswh}. As we said, the dynamical evolution is determined
by Eqs. (\ref{sigma}) and (\ref{pe}), or by any of them and Eq. ({\ref{conserva}), and to complete the system
we must add an equation of state that relates $p$ with $\sigma$.

Our first move to address the stability issue is to recast Eq. (\ref{sigma}) in such a way that it allows us to get $\dot{a}={\cal F}(a, \sigma(a))$. Then, by squaring appropriately the energy density, we obtain a quadratic polynomial in the variable $X=a^{-2}(1+{\dot{a}}^{2})$  as it reads  
\begin{equation}
\label{mastereq}
r^{2}_{~c}X^{2}-2X\Big(r_{c}\bar{\sigma}+8\Big)+{\bar{\sigma}}^{2}-16G(a)=0
\end{equation}
with $\bar{\sigma}=\sigma/3$, and $G(a)=-\mu/a^{4}$. From the master equation (\ref{mastereq}) we get a single  dynamical equation which completely determines the motion of the wormhole throat after the energy density is selected:
\begin{eqnarray}
\label{Fm}
{\dot{a}}^{2}&=&-V(a)\\
\label{potential}
V(a)&=&1-\frac{a^{2}}{r_{c}}\left[\bar{\sigma} +\frac{4}{r_{c}}\Big( 2+\epsilon Y(a)\Big)\right]
\\
\label{fY}
Y(a)&=&\Big(4+r^{2}_{c}G+ r_{c}\bar{\sigma}\Big)^{\frac{1}{2}}
\end{eqnarray}
where $\epsilon$ denotes either $+1$ or $-1$. For $\mu=0$, Eq.(\ref{Fm}) is similar to the Friedmann one  found by Maeda et al in the context of brane world  cosmology with induced gravity \cite{jc4}.
In order to keep the potential defined on a real domain the following reality condition must hold:
\begin{equation}
a^{4}\Big(4+r_{c}\bar{\sigma} \Big)-r^{2}_{c}\mu\geq 0
\end{equation}
Now, making a Taylor expansion to second order of the potential $V$ around the static solution yields
\begin{eqnarray*}
V(a)=V(a_{0})&+& V'(a_{0})(a-a_{0}) + \frac{1}{2}V''(a_{0})(a-a_{0})^{2}\\
&+& {\cal O}[(a-a_{0})^{3}]
\end{eqnarray*}
From Eq.(\ref{potential}) we get that the first derivative of $V$ is
\[
V'=-\frac{2a}{r_{c}}\Big[\bar{\sigma} +\frac{4}{r_{c}}\left( 2+\epsilon Y(a)~ \right)\Big]
- \frac{a^{2}}{r_{c}}\Big[\bar{\sigma}'+ \frac{4\epsilon}{r_{c}}Y'(a) \Big] 
\]
whereas the second derivate  is given by 
\begin{eqnarray*}
V''=&-&\frac{2}{r_{c}}\Big(\bar{\sigma} + \frac{4}{r_{c}}\left( 2+\epsilon Y(a)~ \right)\Big) - \frac{4a}{r_{c}}\Big(\bar{\sigma}'+ \frac{4\epsilon}{r_{c}}Y'(a) \Big)\\
&-&\frac{a^{2}}{r_{c}}\Big(\bar{\sigma}''+ \frac{4\epsilon}{r_{c}}Y''(a)\Big) 
\end{eqnarray*}
Now, it is useful to rewrite the energy conservation as $a\bar{\sigma}'(a)=-3(\bar{p}+\bar{\sigma})$, where $\bar{p}=p/3$. Using the latter identity we can obtain the second derivative of the energy density 
\begin{equation}
\bar{\sigma}''(a)=\frac{3}{a^{2}}(4+3\eta)(\bar{\sigma}+\bar{p})
\end{equation}
where the parameter $\eta$ is defined by the relation 
\begin{equation}
\eta(\sigma)=\frac{\partial p}{\partial \sigma}
\end{equation} 
which for ordinary matter would represent the squared speed sound: $v^{2}_{~s}=\eta$. Here, however, we simply consider $\eta$ as a parameter entering the equations of state. Besides, we shall be interested in the stability of wormholes supported by ordinary matter like those found in the last section. 
To study the stability of the static solutions under perturbations
preserving the spherical symmetry we linearize the equation of state around the static solution as follows
\begin{equation}
p-p_{0}= \eta_{0}(\sigma-\sigma_{0})
\end{equation}
where the surface energy density $\sigma_{0}$ and the transverse pressure $p_{0}$ for a static configuration ($a = a_0$, $\dot{a}=0$,  and $\ddot{a}=0$) are given by
\begin{eqnarray}
\label{sigo} 
\sigma_{0}&=& \frac{3r_{c}}{a^{2}_{0}}-\frac{12}{a_{0}}\sqrt{f(a_{0})},
\\
\label{po} 
p_{0}&=&- \frac{r_{c}}{a^{2}_{0}}+2\Big(\frac{4f(a_{0})+a_{0} f'(a_{0}) } {a_{0}\sqrt{f(a_{0})}} \Big)
\end{eqnarray}
When evaluating the potential at the static solution $a=a_{0}$  it is easy to see that $V(a_{0})=V'(a_{0})=0$, so the potential is 
\begin{equation}
\label{pot}
V(a)= \frac{1}{2}V''(a_{0})(a-a_{0})^{2}+{\cal O}[(a-a_{0})^{3}]
\end{equation}
where the second derivatives has three parts as it can be seen below

\begin{equation}
\label{Vdseg}
V''(a_{0})=V''_{I} (a_{0})+V''_{II}(a_{0}) +V''_{III}(a_{0}),
\end{equation}
with 
\begin{eqnarray}
\label{Vparts}
V''_{I}=-\frac{2}{r_{c}}\Big( \bar{\sigma}_{0} +\frac{4}{r_{c}}\left( 2+\epsilon Y(a_{0})~ \right)\Big),
\\
V''_{II}=-\frac{4a_{0}}{r_{c}}\Big(-\frac{3}{a_{0}}(\bar{\sigma}_{0}+\bar{p}_{0})+ \frac{4\epsilon}{r_{c}}Y'(a_{0}) \Big),
\\
V''_{III}=-\frac{a^{2}_{0}}{r_{c}}\Big( \frac{3}{a^{2}_{~0}}(4+3\eta_{0})(\bar{\sigma}_{0}+\bar{p}_{0})+ \frac{4\epsilon}{r_{c}}Y''(a_{0}) \Big).
\end{eqnarray}

Before proceeding with the stability issue we must verify  that the following constraints  hold at the same time:
\begin{eqnarray}
\label{C1}
&\mbox{I}& ~~\sigma_{0}>0,
\\
\label{C1}
&\mbox{II}& ~\sigma_{0}+p_{0}>0,
\\
\label{C1}
&\mbox{III}&~ a^{4}\Big(4+r_{c}\frac{\sigma_{0}}{3}\Big)-r^{2}_{c}\mu\geq 0.
\end{eqnarray}
These conditions must be supplemented with IV: $a_{0}>\sqrt{\mu}$ to ensure the existence of the  wormholes manifold. Notice that I and II refer to the energy conditions; that is, we are only interested in wormholes supported by ordinary matter, while  inequality III is completely necessary to keep the potential defined on a real domain. From  Fig.6 we obtain that the  space spanned by the above restrictions is quite representative, in the sense that  it covers a vast region. So the wormholes  are physically relevant and do not correspond to a  finely tuned value of mass $\mu$ or crossover scale $r_{c}$. Of course, there are  others types of solutions but they are not as interesting as these. 

\begin{figure}[!h]
\includegraphics[height=5cm, width=6cm]{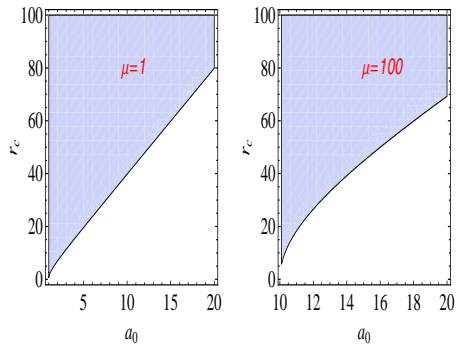}
\caption{We show zones, in the plane $r_{c}-a_{0}$, where the constrains I-IV are satified for $\mu=1,100$.}
\end{figure}
To sum up, wormholes are stable if and only if $V''(a_{0})>0$, while for $V''(a_{0})<0$ perturbations can grow,
 at least until the nonlinear regime is reached. 

From now on, without loss of generality, we consider the stability zones with $\mu=1$. Let us  investigate  how the  regions of stability change with the parameters $r_{c}$ and $\epsilon$. In the case of $\epsilon =-1$ and $r_{c}=10$  we get that there are  stable wormholes with $\eta_{0}\geq 2$  which would correspond to superluminal sound velocity in the wormhole throat (see Fig.7 ). For $\epsilon =\pm 1$ and $r_{c}=100$ wormholes are stable only if $\eta_{0}<0$.  However, for $\epsilon=+1$ and $r_{c}=10$ the model exhibits stable wormholes with $\eta_{0}\geq 0$  only for small radii close to $a_{0}\gtrsim  1$. So, the theory admits classical stable  wormholes with $0\leq\eta_{0}\leq0.1$, indicating that $\eta_{0}$ could represent the speed of sound of nonrelativistic matter (see Fig.7). Interestingly, we get that stable wormholes  are achieved with  values of  the $r_{c}$ parameter far away from the general relativity limit ($r_{c}\rightarrow 0$).
Another novel characteristic introduced by the induced gravity theory called DGP is that these wormholes not only seem to be stable for many different choices of  $\mu$, $r_{c}$, and $\epsilon$, but also they fulfill energy conditions(see Fig.8), being these conditions completely independent of  $\epsilon$. Notice that wormholes fulfilling energy conditions are not possible within Einstein's gravity \cite{visser}. Besides, our results about stable wormholes are in agreement with the stable solutions found in \cite{jc1b} for the full DGP theory  which  supports superluminal excitations.
\begin{figure}[!h]
\includegraphics[height=8cm, width=8cm]{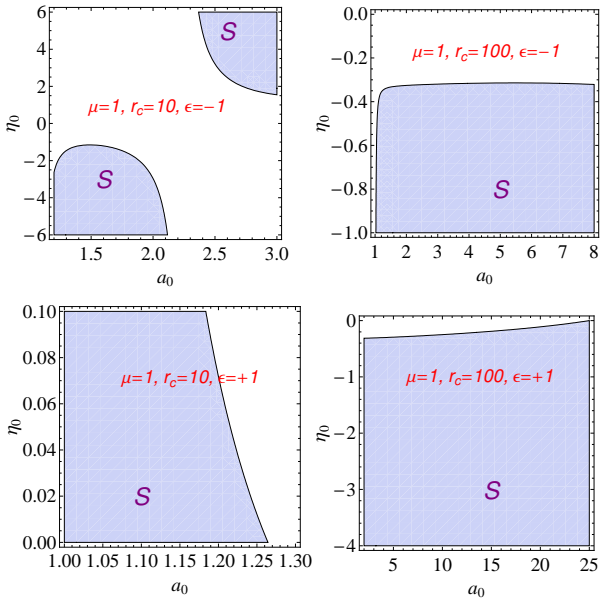}
\caption{We show stability zones  with $\mu=1$,  $r_{c}$=10,100 and for the branches $\epsilon=\pm1$}
\end{figure}
\begin{figure}[!h]
\includegraphics[height=8cm, width=8cm]{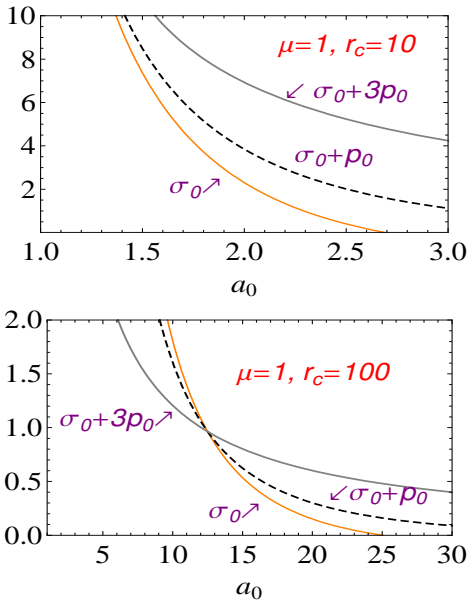}
\caption{We show the energy conditions in terms of $a_{0}$ for $\mu=1$ $r_{c}=10,100$.}
\end{figure}

As a final comment, let us consider the attractive or repulsive character of the wormhole geometry. The wormholes studied could be either attractive or repulsive. To characterize this aspect of the configurations we analyze the force on a test particle at rest in the geometry described above. For this, we evaluate the radial acceleration given by 
\begin{equation}
{\cal A}^{r}=-\Gamma^{r}_{~tt}\Big( \frac{d t}{d \tau}\Big)^{2}
\end{equation}
The sign of the acceleration of a particle initially at rest is then given by  minus the sign of the component $\Gamma^{r}_{~tt}$ of the connection, which for the metric considered is equal to $f'f/2$. Thus we have an attractive gravitational field for $f'>0$ and a repulsive field for $f'<0$ (of course we consider only the possibility $f>0$). In our case, we find that $f'(r)=2\mu/r^{3}$ so the gravitational field turns out to be attractive, indicating that the wormholes are always attractive as long as $\mu>0$. 
\section{Summary}
The generalization of Einstein gravity in the way proposed by  Dvali, Gabadadze, and Porrati introduces a new parameter, which allows for more freedom in the framework of determining the most viable  wormhole configurations. If  wormholes could actually exist, one would be interested in those which  require as little amount of exotic matter as possible. Of course, the case could be that a given change of the theory leads to a worse situation, i. e., that configurations  require more matter violating the energy conditions as the departure from the standard theory becomes relevant. 
However, for suitable wormhole radius, this  seems not to be the case with DGP gravity : Here we have examined the ``exotic'' matter content of thin-shell wormholes using the generalized junction condition, and we  have found that for large values of the DGP parameter, corresponding to a situation far away from the general relativity limit, the amount of exotic matter is reduced in relation to the standard case because it can be positive definite. Moreover, the remarkable result is that we have a region with $\sigma_{0}\geq0$ and besides $\sigma_{0}+ p_{0}\geq0$ , so the WEC and the NEC are satisfied. Further the SEC condition holds also. Thus if the requirement of exotic matter is considered as the hardest objection against wormholes, our results suggest that in a physical scenario with small crossover scale [$r_{c}\sim O(1)$] or far away from the general relativity limit where the DGP becomes dominant  ($r_{c}\gtrsim~10^{2}$) these types of wormholes could be  possible.    
In addition, we showed the existence of  gravitational solitonic wormholes/shell characterized by $\sigma=p=0$  within the DGP model. Unlike the case of Einstein-Gauss-Bonnet theory we found that the existence of  solitonic shells in  DGP gravity  does not  require the presence of a cosmological constant term in the bulk solution.  

Besides, we also focused on the mechanical  stability of  wormhole configurations under radial perturbations preserving the spherical symmetry. We found stable wormholes supported by ordinary matter, that is, configurations which verify the energy conditions for many different choices of the parameter space  spanned by $\mu$, $r_{c}$, and $\epsilon$. Further, the stability analysis shows that in a scenario with a crossover scale $r_{c} \sim {\cal O}(10)$ (far away from the general relativity limit) stable wormholes with very small  squared speed sound $v^{2}_{s} \in [0, 0.1]$ are obtained. Finally, studying the  radial acceleration  experimented by a test particle on the gravitational field we obtained that the latter one turns out to be attractive, indicating that the wormholes are always attractive as long as $\mu>0$.   

\acknowledgments
The author is grateful to the referee for his careful reading of the manuscript.
MGR thanks the University of Buenos Aires for partial support under
Project No.X044.  MGR is also supported by  the Consejo Nacional de Investigaciones Cient\'{\i}ficas y T\'ecnicas (CONICET). 


\end{document}